\documentclass[showpacs,12pt,twocolumn]{iopart}
\usepackage{graphicx}
\usepackage{graphics}
\usepackage{epsfig}
\usepackage{CJK}
\usepackage{color}

\begin{document}

\title{Critical current density and vortex pinning mechanism of Li$_{0.32}$(NH$_{3}$)$_{y}$Fe$_{2}$Te$_{1.2}$Se$_{0.8}$ single crystals}
\author{Shaohua Wang$^{1}$, Shanshan Sun$^{1}$, and Hechang Lei$^{1,*}$}
\address{$^{1}$Department of Physics and Beijing Key Laboratory of Opto-electronic Functional Materials $\&$ Micro-nano Devices, Renmin University of China, Beijing 100872, China}
\ead{hlei@ruc.edu.cn}

\begin{abstract}

We grew Li$_{x}$(NH$_{3}$)$_{y}$Fe$_{2}$Te$_{1.2}$Se$_{0.8}$ single crystals successfully using the low-temperature ammonothermal method and the onset superconducting transition temperature $T_{c}^{\rm{onset}}$ is increased to 21 K when compared to 14 K in the parent compound FeTe$_{0.6}$Se$_{0.4}$. The derived critical current density $J_{c}$ increases remarkably to 2.6$\times$10$^{5}$ A/cm$^{2}$ at 2 K. Further analysis indicates that the dominant pinning mechanism in Li$_{x}$(NH$_{3}$)$_{y}$Fe$_{2}$Te$_{1.2}$Se$_{0.8}$ single crystal is the interaction between vortex and surface-like defects with normal core, by variations in the charge-carrier mean free path $l$ near the defects ($\delta l$ pinning). Moreover, the flux creep is important to the vortex dynamics of this material.

\end{abstract}

\maketitle

\section{Introduction}

The iron-based superconductors (IBSCs) have induced great interest since their discovery almost a decade ago. The family of IBSCs exhibits rather high superconducting transition temperature $T_{c}$, large upper critical field $\mu_{0}H_{c2}$ and critical current density $J_{c}$. These unique properties are important not only for basic sciences but also for practical applications. Among iron-chalcogenide SCs, FeCh (Ch = S, Se, and Te) has nearly isotropic $\mu_{0}H_{c2}$ and rather large $J_{c}$\cite{Lei HC1,S. I. Vedeneev,Sun Y1,Yadav,Sun Y2}, but the relatively low $T_{c}$ limits their applications in some extent. When monovalent metals A (A = K, Rb, Cs, and Tl) are intercalated into FeCh, the $T_{c}$ is raised up to about 32 K with rather high $\mu_{0}H_{c2}$ ($\sim $ 56 T for $H\Vert c$ at 1.6 K)\cite{Guo,Mun}. However, for A$_{x}$Fe$_{2-y}$Se$_{2}$, there are Fe vacancies in the FeCh layer\cite{Bao}. More severely, the superconducting phase always intergrows with the insulating phase A$_{0.8}$Fe$_{1.6}$Se$_{2}$, leading to a mesoscopic phase separation\cite{F Chen}. Correspondingly, the superconducting phase takes over only small parts of the total phase. On the one hand, that impedes the investigation of intrinsic superconducting properties of these materials. On the other hand, it also results in the rather small $J_{c}$ of A$_{x}$Fe$_{2-y}$Se$_{2}$ even compared to FeCh\cite{Lei HC2,Li MT}.

Recently, superconductivity with $T_{c}$ up to about 45 K has been reported in AM$_{x}$(NH$_{3}$)$_{y}$Fe$_{2}$Ch$_{2}$ (AM = alkali, alkali-earth, and rare-earth metals)\cite{Ying,Scheidt,Burrard,Ying2,Sedlnaier,Guo JG,Lei HC3}. Previous studies indicate that the Fe vacancies are almost absent in these materials\cite{Ying2,Lei HC3}. Thus, it is promising that AM$_{x}$(NH$_{3}$)$_{y}$Fe$_{2}$Ch$_{2}$ will have a relative high $J_{c}$. Moreover, the enhanced $T_{c}$ and mass anisotropy due to large interlayer distance along the $c$ axis in AM$_{x}$(NH$_{3}$)$_{y}$Fe$_{2}$Ch$_{2}$ could result in the significant increase of Ginzburg number $G_{i}$, i.e., the vortex motion and fluctuations would become quite strong. It causes some very interesting phenomena in vortex dynamics, such as giant-flux creep and thermally activated flux flow etc. Because of the difficulty of single crystal growth for AM$_{x}$(NH$_{3}$)$_{y}$Fe$_{2}$Se$_{2}$, related study is still absent.

In this work, we report the study on the $J_{c}$ of Li$_{x}$(NH$_{3}$)$_{y}$Fe$_{2}$Te$_{1.2}$Se$_{0.8}$ (LiFeTeSe-122) single crystals synthesized by the low-temperature ammonothermal method. The $J_{c}$ reaches 2.6$\times$10$^{5}$ A/cm$^{2}$ at 2 K. The detailed analysis suggests that the main pinning sources are surface-like defects with normal core and the flux creep can not be ignored when analysing vortex dynamics of this material.

\section{Experimental}

The FeTe$_{0.6}$Se$_{0.4}$ single crystals were grown by self-flux method with nominal ratio of Fe : Te : Se = 1 : 0.6 : 0.4. Fe pieces (99.98 \%), selenium shots (99.999 \%), and Te grains (99.99 \%) were mixed and loaded into alumina crucible, which was sealed in the quartz tube under partial argon atmosphere. The sealed ampoule was heated to 1273 K and kept at this temperature for 24 h. Then it was cooled to room temperature slowly. The LiFeTeSe-122 single crystals were synthesized by the low-temperature ammonothermal technique\cite{Ying,Scheidt,Burrard,Ying TP}. The pieces of Li metal and FeTe$_{0.6}$Se$_{0.4}$ single crystals in the molar ratio of 1 : 2 were loaded into the high-pressure vessel (25 mL) with a magnetic stirrer. All of these processes were carried out in an argon-filled glovebox with O$_{2}$ and H$_{2}$O content below 0.1 ppm. Then, the vessel was taken out from glovebox and connected to a vacuum line equipped with a molecular pump and a NH$_{3}$ gas line. The vessel was evacuated by using a molecular pump ($\sim$ 1$\times$10$^{-3}$ Pa) before introducing NH$_{3}$ and placed in an ethanol bath cooled to $\sim$ 238 K, then the NH$_{3}$ gas was condensed into the vessel for 20 minutes. After that, the vessel was taken out from the cooling bath and stirred for 2 days at room temperature in order to facilitate the reaction and to improve the homogeneity of intercalation. Finally, the NH$_{3}$ gas was evacuated using a molecular pump. X-ray diffraction (XRD) patterns were collected using a Bruker D8 X-ray Diffractometer with Cu $K_{\alpha}$ radiation ($\lambda=$ 0.15418 nm) at room temperature. Rietveld refinement of the XRD patterns was performed using the code TOPAS4\cite{TOPAS}. The elemental analysis was performed using the inductively coupled plasma atomic emission spectroscopy (ICP-AES). Magnetization measurements were performed in a Quantum Design Magnetic Property Measurement System (MPMS3) up to 5 T.

\section{Results and discussion}

\begin{figure}[tbp]
\centerline{\includegraphics[scale=0.6]{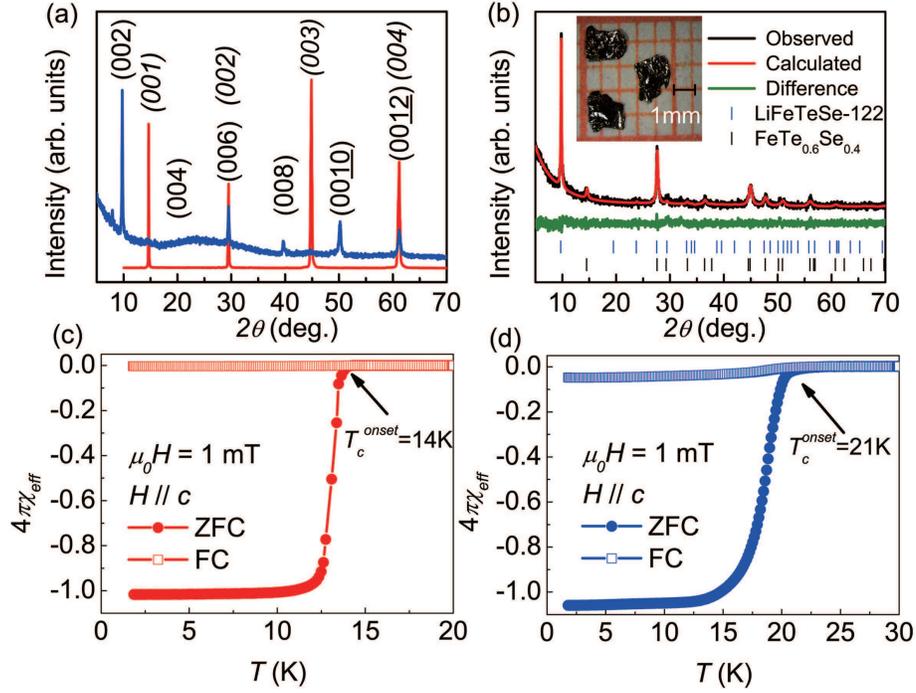}} \vspace*{-0.3cm}
\caption{(a) XRD patterns of FeTe$_{0.6}$Se$_{0.4}$ and  LiFeTeSe-122 single crystals shown in red and blue. (b) Powder XRD pattern of LiFeTeSe-122 and the fitting result when considering two phases of LiFeTeSe-122 and FeTe$_{0.6}$Se$_{0.4}$. Inset: photograph of typical LiFeTeSe-122 crystals on 1 mm grid paper. (c) and (d) Temperature dependence of magnetic susceptibility 4$\pi$$\chi_{eff}(T)$ at low temperature region for FeTe$_{0.6}$Se$_{0.4}$ and LiFeTeSe-122 single crystals with zero-field-cooling and field-cooling modes ($\mu_{0}H=$ 1 mT, $H\Vert c$).}
\end{figure}

Fig. 1(a) shows the XRD patterns of FeTe$_{0.6}$Se$_{0.4}$ and LiFeTeSe-122 single crystals. Only $(00l)$ reflections can be indexed, indicating that the surfaces of crystals are parallel to the $(00l)$-plane. The diffraction peaks of LiFeTeSe-122 shift to lower angle when compared to FeTe$_{0.6}$Se$_{0.4}$, suggesting the larger interlayer distance in the former. The typical size of LiFeTeSe-122 single crystals is about 1$\times$2 mm$^{2}$ (inset of Fig. 1(b)), similar to the size of parent crystals, i.e., the shape of crystal is roughly unchanged during intercalation process. It notes that the intensity of XRD diffraction peaks of LiFeTeSe-122 is weaker than that of FeTe$_{0.6}$Se$_{0.4}$, leading to more obvious background signal in the former. The weaken diffraction intensity could be due to the increased roughness of the surface of crystals after intercalation. Fig. 1(b) shows the powder XRD pattern of LiFeTeSe-122 and the fitted $a-$ and $c$-axial lattice parameters are 3.8270(8) and 18.17(1) \AA\, consistent with the previous results \cite{Lei HC3}. There are weak diffraction peaks originating from Fe(Te, Se). It is not due to the incomplete intercalation of Li-NH$_{3}$ but the decomposition of LiFeTeSe-122 during grinding for powder XRD measurement. If the intercalation is incomplete, the superconducting transition of Fe(Te, Se) with $T_{c}\sim$ 15 K would be clearly observed in the curves of magnetic susceptibility, which is not the case in our experiment (shown below). The atomic ratio of Li : Fe : Te : Se determined from the ICP-AES analysis is 0.16 : 1 : 0.60 : 0.38. The molar ratio of Te to Se is perfectly consistent with the nominal ratio of parent compound. More importantly, there is no Fe vacancy in the LiFeTeSe-122 crystals. Temperature dependence of magnetic susceptibility 4$\pi\chi_{eff}(T)$ at low temperature region for $H\Vert c$ (Fig. 1(c) and (d)) clearly shows the superconducting transition in both samples. The onset superconducting transition temperature $T_{c}^{\rm{onset}}\sim$ 14 K for FeTe$_{0.6}$Se$_{0.4}$, consistent with previous results in the literature\cite{Lei HC3}. After intercalation, the $T_{c}^{\rm{onset}}$ is enhanced to about 21 K, which is slightly higher than that of powder sample\cite{Lei HC3}.
After considering the demagnetization effect of sample by using the formula $4\pi\chi_{eff}=4\pi\chi/(1-4\pi N_{d}\chi)$ where $N_{d}$ is demagnetization factor \cite{Aharoni} (0.70 and 0.82 for FeTe$_{0.6}$Se$_{0.4}$ and LiFeTeSe-122, respectively), the estimated superconducting volume fractions (SVFs) from zero-field-cooling 4$\pi\chi_{eff}(T)$ curves at 2 K for both samples are $\sim$ 100 \%, clearly indicating the bulk superconductivity in these crystals. On the other hand, the smaller SVFs determined from the field-cooling curves imply that both of them are type-II superconductors with rather strong vortex pinning effects.

\begin{figure}[tbp]
\centerline{\includegraphics[scale=0.35]{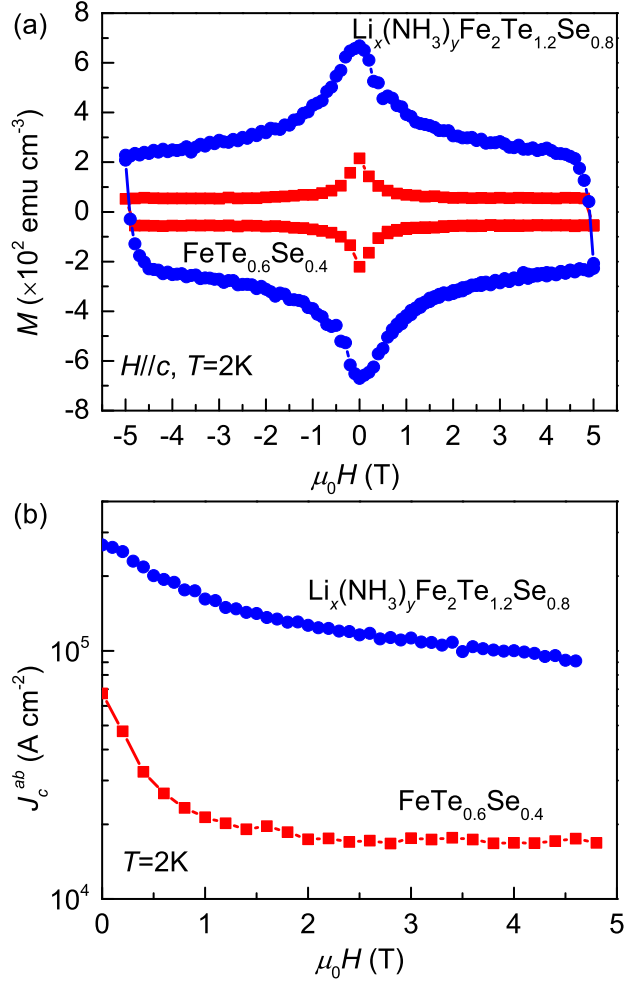}} \vspace*{-0.3cm}
\caption{(a) Magnetization hysteresis loops of LiFeTeSe-122 and FeTe$_{0.6}$Se$_{0.4}$ single crystals at 2 K for $H\Vert c$. (b) The derived critical current density $J_{c}^{ab}(\mu _{0}H)$ for two samples at $T=$ 2 K.  }
\end{figure}

The magnetization hysteresis loops (MHLs) of LiFeTeSe-122 and FeTe$_{0.6}$Se$_{0.4}$ single crystals at $T=$ 2 K for $H\Vert c$ are shown in Fig. 2(a). The symmetrical shapes of MHLs for two samples are typical of type-II superconductors, indicating that the bulk pinning is dominant without ferromagnetic impurity in the samples. Importantly, the MHL of the intercalated crystal is much larger than that of parent one, implying that the pinning force is greatly enhanced in the intercalated sample. According to the Bean model\cite{Bean,E.M.}, the critical current density can be determined from the MHLs. For a rectangularly-shaped crystal with dimension $c<a<b$, when $H\Vert c$, the in-plane critical current density $J_{c}^{ab}(\mu _{0}H)$ is given by

\begin{equation}
J_{c}^{ab}(\mu _{0}H)=\frac{20\Delta M(\mu _{0}H)}{a(1-a/3b)}
\end{equation}

where $a$ and $b$ ($a<b$) are the in-plane sample size in cm, $\Delta M(\mu_{0}H)$ is the difference between the magnetization values for increasing and decreasing fields at a particular applied field value (measured in emu/cm$^{3}$), and $J_{c}^{ab}(\mu _{0}H)$ is the critical current density in A/cm$^{2}$. As shown in Fig. 2(b), at $T=$ 2 K, the $J_{c}^{ab}(\mu _{0}H)$ of FeTe$_{0.6}$Se$_{0.4}$ at self field is about 6.7$\times$10$^{4}$ A/cm$^{2}$. In contrast, the $J_{c}^{ab}(\mu _{0}H)$ of LiFeTeSe-122 at self-field is about 4 times larger than that of parent sample and reaches 2.6$\times$10$^{5}$ A/cm$^{2}$. Moreover, the decrease of $J_{c}^{ab}(\mu _{0}H)$ is rather slow with increasing magnetic field, suggesting the strong vortex pinning effect in the sample. It has to be mentioned that because of the bulk superconductivity with large SVF in LiFeTeSe-122, its $J_{c}^{ab}$ is even much larger than that of K$_{x}$Fe$_{2-y}$Se$_{2}$ which has much higher $T_{c} (\sim$ 32 K)\cite{Lei HC4,Gao}. However, the $J_{c}^{ab}$ of LiFeTeSe-122 is still low when compared to the iron pnictide superconductors where the typical self-field $J_{c0}^{ab}$ is above 10$^{6}$ A/cm$^{2}$ at 2 K\cite{Yang H}.

\begin{figure}[tbp]
\centerline{\includegraphics[scale=0.35]{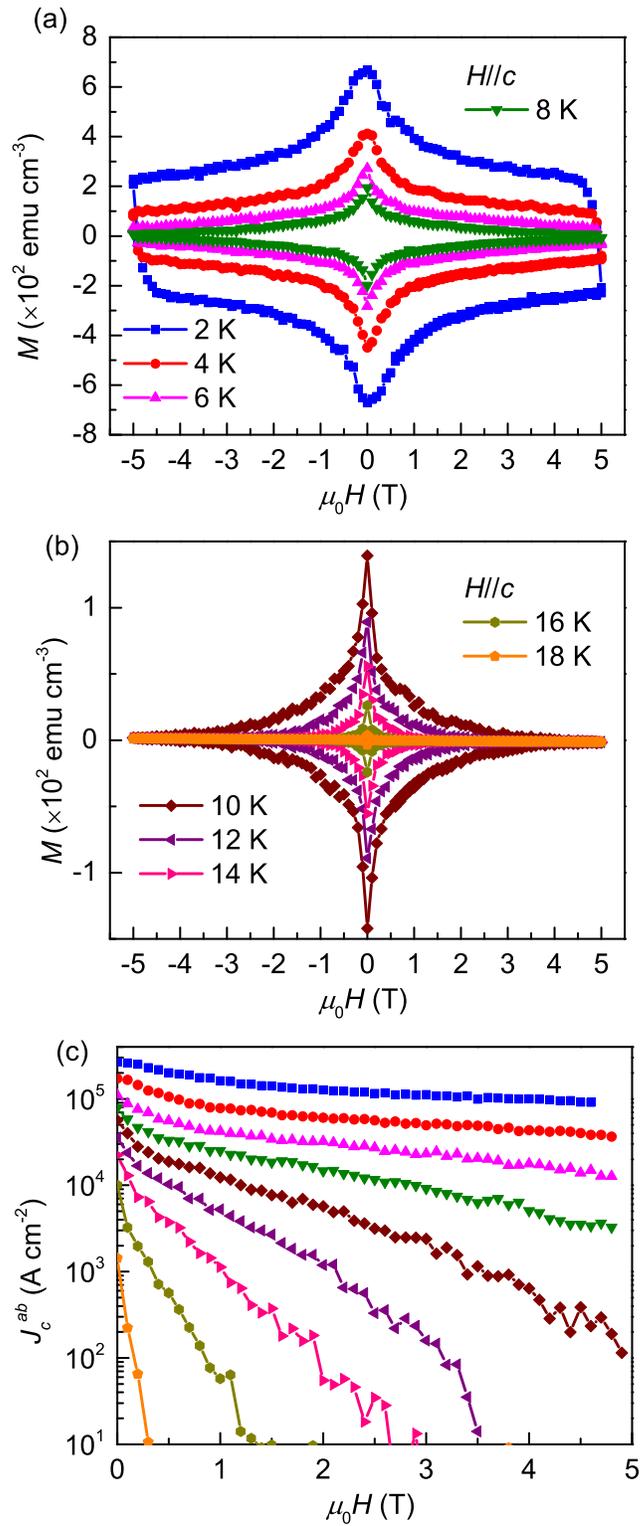}} \vspace*{-0.3cm}
\caption{(a) The MHLs of LiFeTeSe-122 single crystal at the temperatures range of 2 K to 8 K and (b) 10 K to 18 K. (b) The $J_{c}^{ab}(\mu _{0}H)$ at corresponding temperatures derived from (a) using the Bean model.}
\end{figure}

Fig. 3(a) and (b) show the MHLs of LiFeTeSe-122 single crystal for $H\Vert c$ at the temperatures range of 2 K to 8 K and 10 K to 18 K, respectively. The hysteresis area decreases with increasing temperature, indicating the $J_{c}^{ab}$ decreases as temperature increasing. The derived $J_{c}^{ab}(\mu _{0}H)$ of LiFeTeSe-122 single crystal at different temperatures from the MHLs using eq. (1) is shown in Fig. 3(b). The $J_{c}^{ab}(\mu _{0}H)$ are robust against the applied field at low temperatures, but the slopes of $J_{c}^{ab}(\mu _{0}H)$ vs. $\mu_{0}H$ become larger at high temperatures, indicating a significant thermally-activated depinning process.

\begin{figure}[tbp]
\centerline{\includegraphics[scale=0.4]{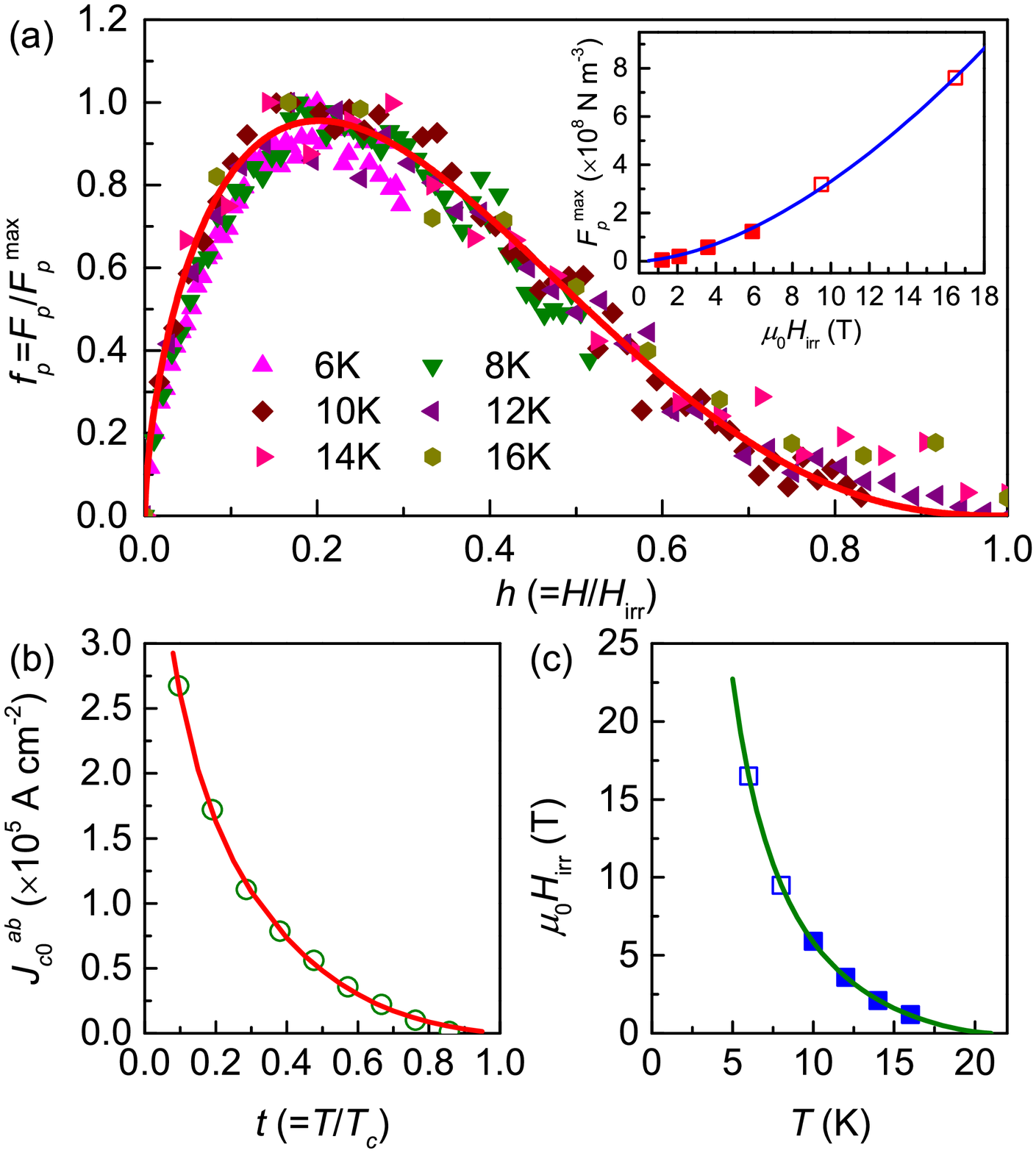}} \vspace*{-0.3cm}
\caption{(a) Normalized vortex pinning force $f_{p}=F_{p}/F_{p}^{\rm{max}}$ as a function of normalized field $h=H/H_{\rm{irr}}$ for LiFeTeSe-122 single crystal at various temperatures. Solid line represents the fitting curve using $f_{p}\propto h^{p}(1-h)^{q}$. Inset shows the $F_{p}^{\rm{max}}$ as a function of $\mu _{0}H_{\rm{irr}}$. The fitting result using $F_{p}^{\rm{max}}=A(\mu _{0}H_{\rm{irr}})^{\alpha}$ is shown as solid line. The measured and estimated $\mu _{0}H_{\rm{irr}}$ are shown as closed and open circles. (b) Reduced temperature dependence of self-field $J_{c0}^{ab}(t)$. The solid line is the theoretical curve obtained based on eq. (2) with the coexistence of $\delta l$ and $\delta T_{c}$ pinning mechanisms. (c) Temperature dependence of $\mu _{0}H_{\rm{irr}}(T)$. Solid line is the fitting using the flux creep model.}
\end{figure}

The vortex pinning force $F_{p} (= \mu_{0}H\times J_{c}^{ab}$) can provide more information about the vortex pinning mechanism in LiFeTeSe-122 single crystal. According to the Dew-Huges model\cite{Dew-Hughes}, if one pinning mechanism is dominant in certain temperature range, the normalized vortex pinning force $f_{p} = F_{p}/F_{p}^{\rm{max}}$ at different temperatures should be proportional to $h^{p}(1-h)^{q}$, where $F_{p}^{\rm{max}}$ is the maximum pinning force, the indices $p$ and $q$ are determined by the pinning mechanism, $h = H/H_{\rm{irr}}$ is the normalized field, and the irreversibility field $\mu _{0}H_{\rm{irr}}$ is estimated by extrapolating $J_{c}^{ab}(T,\mu _{0}H)$ to zero. Fig. 4(a) shows the relationship between $f_{p}(h)$ and $h$ at different temperatures ($T>$ 6 K) for $H\Vert c$. It can be clearly seen that the $f_{p}(h)$ as a function of $h$ exhibits a temperature independence scaling law, suggesting the dominance of single pinning mechanism. The fitting using $f_{p}(h)\propto h^{p}(1-h)^{q}$ gives $p = 0.63(3)$ and $q = 2.52(8)$. The value of $h_{\rm{fit}}^{\rm{max}}$ calculated by $p/(p+q)$ equals 0.202, consistent with the peak positions obtained from the experimental curves at all temperatures $h_{\rm{exp}}^{\rm{max}}\approx$ 0.207. Moreover, for $T=$ 6 and 8 K, the $H_{\rm{irr}}$ could be estimated by locating the field of $F_{p}^{\rm{max}}$ at $h_{\rm{exp}}^{\rm{max}}$. Partial $f_{p}(h,T)$ curves at $T=$ 6 and 8 K also exhibit the same scaling law, suggesting that the same pinning mechanism is dominant above 6 K. On the other hand, when $T<$ 6 K, the scaling behavior cannot be analyzed because of the absence of $F_{p}^{\rm{max}}$. The values of $p$, $q$ and $h_{\rm{fit}}^{\rm{max}}$ are close to the expected values ($p=$ 0.5, $q=$ 2, and $h^{\rm{max}}=$ 0.2) for the pinning of surface-like defects with normal core\cite{Dew-Hughes}. The slightly larger values of $p$ and $q$ than theoretical predictions suggest that the flux creep might have some influences on the pinning force\cite{T. Matsushita}. Interestingly, similar values of $p$ and $q$ have also been observed in FeS and FeS$_{0.94}$Se$_{0.06}$ single crystals prepared by deintercalating potassium from K$_{x}$Fe$_{2-y}$(S, Se)$_{2}$ using the hydrothermal method\cite{Wang AF}. It suggests that there is a common type of pinning centers in these materials in contrast to FeSe$_{0.5}$Te$_{0.5}$ thin films, K$_{x}$Fe$_{2-y}$Se$_{2}$, and Ba$_{0.6}$K$_{0.4}$Fe$_{2}$As$_{2}$ where the point-like defects with normal core are the dominant pinning sources\cite{Lei HC4,Yang H,Yuan PS}. Moreover, the $F_{p}^{\rm{max}}$ can be fitted using $F_{p}^{\rm{max}}=A(\mu _{0}H_{\rm{irr}})^{\alpha}$ (inset of Fig. 4(a)) and the obtained $\alpha$ is 1.67(4), also close to the theoretical value ($\alpha$ = 2)\cite{Dew-Hughes}.

On the other hand, the self-field $J_{c0}^{ab}$ reduces quickly at low-temperature region and then this trend becomes milder at higher temperatures (Fig. 4(b)). It implies that flux creep needs to be considered in vortex dynamics of LiFeTeSe-122 single crystals. In the framework of the thermally-activated flux motion model considering collective flux pinning and creep effect\cite{L Civale,S. R. Ghorbani} the $J_{c}^{ab}(t)$ can be expressed as

\begin{equation}
J_{c}^{ab}(t)=\frac{J_{c0}(0)J(t)}{\{1+[\mu CT_{c}t/g(t)]\}^{1/\mu}}
\end{equation}

where $t=T/T_{c}$ is the reduced temperature, $C$ is a temperature-independent constant, $J_{c0}(0)$ is the critical current density at $t=$ 0, $J(t)$ and $g(t)$ is the temperature-dependent parts of critical current density and characteristic pinning potential when the flux creep is absent, respectively, $\mu$ is the glassy exponent, related to the size of the vortex bundle in the collective creep theory, and in a three-dimensional system, it is predicted to be 1/7, 2/3, and 7/9 for single-vortex, small-bundle, and large-bundle regimes, respectively\cite{M. V. Feigel'man}. According to the collective theory, there are two pinning mechanisms $\delta T_{c}$ and $\delta l$, related to the spatial variations of $T_{c}$ and charge-carrier mean free path $l$ near defects, respectively. For both pinning mechanisms, the temperature dependence of the $J(t)$ and $g(t)$ is different. For $\delta T_{c}$ pinning, $J^{\delta T_{c}}(t)=(1-t^{2})^{7/6}(1+t^{2})^{5/6}$ and $g^{\delta T_{c}}(t)=(1-t^{2})^{1/3}(1+t^{2})^{5/3}$ while for $\delta l$ pinning, $J^{\delta l}(t)=(1-t^{2})^{5/2}(1+t^{2})^{-1/2}$ and $g^{\delta l}(t)=1-t^{4}$\cite{R. Griessen}. Assuming the coexistence of $\delta T_{c}$\ and $\delta l$ pinning mechanisms, we have $J_{c}^{ab}(t)=xJ_{c}^{ab,\delta T_{c}}(t)+(1-x)J_{c}^{ab,\delta l}(t)$, where $x$ represents the contribution of $J_{c}^{ab,\delta T_{c}}(t)$. The $J_{c0}^{ab}(t)$ can be well fitted using eq. (2) (Fig. 4(b)). The fitted $x$ is 0.15(5) with $\mu=$ 1.1(1). The value of $\mu$ is between the prediction of small-bundle and large-bundle regimes and similar to the value observed in FeTe$_{0.6}$Se$_{0.4}$ single crystals\cite{Sun Y}. The small but non-zero $x$ indicates that both $\delta T_{c}$\ and $\delta l$ pinning mechanisms play roles in LiFeTeSe-122 single crystals, but the latter one is dominant.

According to the theory of thermally-activated flux creep and assuming $\mu_{0}H_{\rm{irr}}$ is small when compared to $\mu_{0}H_{c2}$\cite{T. Matsushita2,G. Fuchs}, the temperature dependence of $\mu_{0}H_{\rm{irr}}(T)$ can be described with three parameters $K$, $m$, and $\gamma$ as

\begin{equation}
\mu_{0}H_{\rm{irr}}(T)=\left(\frac{K}{T}\right)^{4/(3-2\gamma)}\left[1-\left(\frac{T}{T_{c}}\right)^{2}\right]^{2m/(3-2\gamma)}
\end{equation}

As shown in Fig. 4(c), the temperature dependence of $\mu_{0}H_{\rm{irr}}(T)$ agrees well with the theoretical fitting based on the flux creep model and the obtained parameters are $K=$ 40(1), $m=$ 1.9(1), and $\gamma=$ 0.20(3). The values of $m$ and $\gamma$ are similar to those in Hg-based cuprate superconductors and SmFeAsO$_{0.85}$\cite{G. Fuchs,D. Ahmad}.

Finally, it has to be mentioned that the $J_{c}$ of FeTe$_{0.6}$Se$_{0.4}$ single crystals studied in present work is not the highest one when compared to those reported in the literature \cite{Sun Y}. It can be due to different conditions of crystal growth and a number of the defects in the crystals. If the quality of parent compounds Fe(Te, Se) can be improved further, the $J_{c}$ of LiFeTeSe-122 could be even higher. On the other hand, the Fe(Te, Se) films can have larger $J_{c}$ than single crystals because of various kinds of external factors, such as interface effects, nonmagnetic/magnetic point/nanorod defects/inclusions introduced during preparation process of films, pinning of grain boundaries etc \cite{Li,Si,Haindl S,Ozaki}. Moreover, the enhancement of $T_{c}$ has been observed in Fe(Te, Se) films and it is inextricably linked to the strain induced during the epitaxial growth \cite{Si,Bellingeri}. It would be very interesting to examine whether the $J_{c}$ and $T_{c}$ of films could increase further when intercalating Li-NH$_{3}$ with proper doping level of electron carriers.

\section{Conclusion}

In summary, we investigate the critical current density $J_{c}$ of LiFeTeSe-122 single crystals grown using the low-temperature ammonothermal method. The cointercalation of Li and NH$_{3}$ not only increases the $T_{c}$ from 14 K to 21 K, but also significantly increases the $J_{c}$ to 2.6$\times$10$^{5}$ A/cm$^{2}$ at 2 K. Detailed analysis of the vortex dynamics indicates that the dominant pinning sources are surface-like defects with normal core. Moreover, the flux creep is important to the vortex dynamics of LiFeTeSe-122 single crystals and the analysis of self-field $J_{c0}^{ab}$ suggests that the $\delta l$ pinning mechanism due to spatial fluctuations of the charge-carrier mean free path is dominant at measured temperature range.

\ack{
This work was supported by the Ministry of Science and Technology of China (2016YFA0300504), the National Natural Science Foundation of China (No. 11574394), the Fundamental Research Funds for the Central Universities, and the Research Funds of Renmin University of China (RUC) (15XNLF06, 15XNLQ07).}

\end{document}